\documentclass[prc,twocolumn,superscriptaddress,showpacs,preprintnumbers,amsmath,amssymb]{revtex4}
\usepackage[dvips]{graphicx}
\usepackage{dcolumn}
\usepackage[usenames]{color}

\def\m@thcombine#1#2{%
  \setbox0=\hbox{$#1$}
  \setbox1=\hbox{$#2$}
  \ifdim\wd0>\wd1
    \setbox0=\hbox to\wd1{\hss\box0\hss}
  \else
    \setbox1=\hbox to\wd0{\hss\box1\hss}
  \fi
  \mathop{\vcenter{
    \offinterlineskip\box0\box1}}}
\def\lesim{\m@thcombine<\sim}
\def\gesim{\m@thcombine>\sim}

\begin{document}
\title{Low-energy $E1$ strength in select nuclei:
Possible constraints on the neutron skins and the symmetry energy
}
\author{Tsunenori Inakura}
\affiliation{Department of Physics, Graduate School of Science, Chiba University, 
Yayoi-cho 1-33, Inage, Chiba 263-8522, Japan}
\affiliation{RIKEN Nishina Center, Wako, 351-0198, Japan}
\author{Takashi Nakatsukasa}
\affiliation{RIKEN Nishina Center, Wako, 351-0198, Japan}
\affiliation{Center for Computational Sciences, University of Tsukuba, Tsukuba 305-8571, Japan}
\author{Kazuhiro Yabana}
\affiliation{Center for Computational Sciences, University of Tsukuba, Tsukuba 305-8571, Japan}
\affiliation{RIKEN Nishina Center, Wako, 351-0198, Japan}

\begin{abstract}
Correlations between low-lying electric dipole ($E1$) strength 
and neutron skin thickness are systematically investigated
with the
fully self-consistent random-phase approximation using the Skyrme energy
functionals.
The presence of strong correlation among these quantities is currently
under dispute.
We find that the strong correlation is present in properly selected nuclei,
namely in spherical neutron-rich nuclei in the region
where the neutron Fermi levels are located at orbits with
low orbital angular momenta.
The significant correlation between
the fraction of the energy-weighted sum value and the slope
of the symmetry energy is also observed.
The deformation in the ground state seems to weaken
the correlation.
\end{abstract}

\maketitle


The isospin-dependent part of the nuclear equation of state (EOS),
especially the symmetry energy, is receiving current attention
\cite{Ste05,Li08}.
Although the symmetry energy at the saturation density
$E_{\rm sym}(\rho_0)$ is relatively well known,
its values at other densities, which have a strong impact on the
description of neutron stars and steller explosions,
are poorly determined at present.
Information on the density dependence of the symmetry energy might
be obtained from the neutron-skin thickness $\Delta r_{np}$,
since the skin thickness was found to be strongly correlated with the slope $L$
of the symmetry energy; $L=3\rho_0 E'_{\rm sym}(\rho_0)$ \cite{Bro00,Fur02}.
However, the large uncertainties in measured neutron-skin thickness
have practically prohibited us from making an accurate estimate on $L$.

The electric dipole ($E1$) response is a fundamental tool to
probe the isovector property of nuclei.
The giant dipole resonance (GDR), which is rather insensitive to the
structure of an individual nucleus, provides information on the magnitude
of the symmetry energy near the saturation density $\rho_0$.
In contrast, the low-energy $E1$ modes, which are often referred to as pygmy 
dipole resonances (PDR), is sensitive to the nuclear structure, such as
the existence of loosely bound nucleons.
Thus, the PDR, which is currently of significant interest
in physics of exotic nuclei,
may carry information on the symmetry energy $E_{\rm sym}(\rho)$
at densities away from $\rho_0$.

Among many issues on the PDR,
the correlation between the PDR and neutron skin
is one of important subjects currently under dispute.
If the strong correlation exists,
the PDR may constrain both $\Delta r_{np}$ and the slope parameter $L$.
The calculation by Piekarewicz with the random-phase approximation (RPA)
based on the relativistic mean-field model
predicted a linear correlation for Sn isotopes \cite{Piekarewicz06}.
Utilizing similar arguments, the neutron skin thickness and the slope parameter
were estimated from available data in $^{208}$Pb, $^{68}$Ni, 
$^{132}$Sn, and so on \cite{Klimkiewicz,Carbone}.
However, Reinhard and Nazarewicz
performed a covariance analysis investigating the parameter dependence
for the Skyrme functional models,
which concluded that the correlation between the PDR strength and
$\Delta r_{np}$ is very weak \cite{Reinhard}.
Recently, they have extended their studies
to the $E1$ strength at finite momentum transfer $q$ \cite{Reinhard2}.
It should be noted that these conclusions, which seemed to contradict
to each other,
were given from RPA calculations for specific spherical nuclei
using different ways of analysis.

Recently, we have performed a systematic RPA calculation on the
PDR for even-even nuclei \cite{Inakura11}
using the finite amplitude method \cite{Nakatsukasa07,Inakura09,Avogadro11,Stoitsov11,Avogadro13}.
The calculation is self-consistent with the Skyrme energy functional
and fully takes into account the deformation effects.
We found that the significant enhancement of
the PDR strength takes place in regions of specific neutron numbers.
The main purpose of the present paper is to show that
the quality of the correlation between the PDR strength
and $\Delta r_{np}$ are also sensitive to the neutron number of the isotopes.
Namely, the strong correlation exists only in particular neutron-rich nuclei.
This may provide a possible suggestion for future measurements
to constrain $\Delta r_{np}$ and $L$.

{\it Numerical calculations ---}
We perform an analysis similar to Ref.~\cite{Reinhard}
to investigate the Skyrme parameter dependence of the RPA results
for nuclei of many kinds (mostly with $Z\leq 40$), including
stable, neutron-rich, spherical, and deformed nuclei.
The fully self-consistent RPA equation is solved
using a revised version of the RPA code in Ref. \cite{Inakura06}.
The size of the RPA matrix is reduced by
assuming the reflection symmetry of the ground state
with respect to $x=0$, $y=0$, and $z=0$ planes.
We adopt the representation of the three-dimensional adaptive Cartesian 
grids \cite{Nakatsukasa05} within a sphere of the radius
$R_\mathrm{box}=$ 15 fm.
The real-space representation has an advantage over 
other representations, such as harmonic oscillator basis, on the treatment of 
the continuum states. 
The Skyrme functional of the SkM$^\ast$ parameter set \cite{SkM*} is 
used unless otherwise specified.
The residual interaction in the present calculation contains all 
terms of the Skyrme interaction including
the residual spin-orbit, the residual Coulomb, and the time-odd components.
The pairing correlation is neglected for simplicity, which has little impact
on $E1$ modes \cite{Ebata10}.

\begin{figure}[tb]
\begin{center}
\includegraphics[width=0.40\textwidth,keepaspectratio]{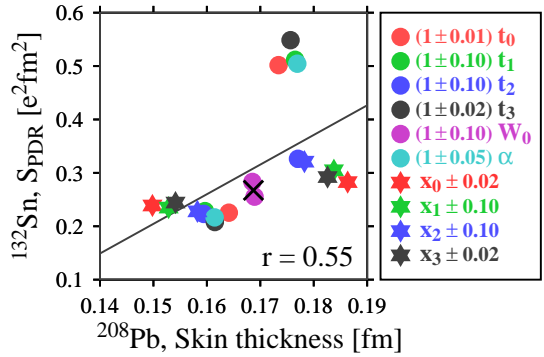}
\caption{(Color online) 
Correlations between
the PDR strength $S_\mathrm{PDR}$ in $^{132}$Sn
and the neutron skin thickness $\Delta r_{np}$ in $^{208}$Pb.
The cross denotes a result obtained 
with the original SkM$^\ast$ parameter set.
Other symbols represent results with the modified parameter set
as shown in the right panel. The solid line indicates a linear fit.
The correlation coefficient for these parameter set is also shown. 
See the text for detail.
}
\label{fig: Pb-Sn}
\end{center}
\end{figure}

{\it Definition of PDR strength, PDR fraction, and correlation coefficient---}
We define the PDR strength as
\begin{equation}
\label{SPDR}
S_\mathrm{PDR}\equiv
 \int_0^{\omega_c} S(E1;E) dE
 =\sum_n^{E_n<\omega_c} B(E1;n) ,
\end{equation}
with the PDR cutoff energy $\omega_c$.
The PDR fraction $f_\mathrm{PDR}$
is the ratio of the integrated photoabsorption cross
section below $\omega_c$ to the total integrated cross section.
\begin{equation}
\label{fPDR}
 f_{\mbox{\sc pdr}}
= \frac{\int^{\omega_c} \sigma_{\rm abs}(E) dE}
                                    {\int \sigma_{\rm abs}(E) dE}
 = \frac{\sum_{n}^{E_n<\omega_c} E_n B(E1;n)}{\sum_n E_n B(E1;n)} ,
\end{equation}
In Eqs. (\ref{SPDR}) and (\ref{fPDR}),
we fix the cutoff at $\omega_c=10$ MeV.
Many former works adopted the same definition \cite{Reinhard,Inakura11},
because of its simplicity.
In light spherical neutron-rich nuclei, the value of $\omega_c=10$ MeV can
reasonably separate the PDR peaks from the GDR.
However, for heavier nuclei, the separation becomes more ambiguous.
It is especially difficult for deformed nuclei.
Later, we introduce another definition of the PDR strength using a variable
$\omega_c$, to check the validity.

To quantify the correlation between two quantities,
we use the correlation coefficient $r$.
When we have data points for $(x_i,y_i)$ with $i=1,\cdots, N_d$,
it is defined by
\begin{equation}
\label{r}
r \equiv \frac{\sum_{i=1}^{N_d} (x_i-\bar{x})(y_i-\bar{y})}
 {\sqrt{\sum_{i=1}^{N_d} (x_i-\bar{x})^2}\sqrt{\sum_{j=1}^{N_d} (y_j-\bar{y})^2}} ,
\end{equation}
where $\bar{x}$ and $\bar{y}$ are the mean values of $x_i$ and $y_i$,
respectively.
The absolute value of $r$ does not exceed the unity.
A perfect linear correlation, $y_i=ax_i+b$,
corresponds to $r=\pm 1$ with
the same sign as that of parameter $a$.
In the followings, the correlation with $r>0$ ($r<0$) is referred to as
``positive'' (``negative'') correlation.

{\it Neutron skin thickness in $^{208}$Pb ---}
First, we confirm the result in Ref. \cite{Reinhard}.
Reference \cite{Reinhard} reported that
the $S_\mathrm{PDR}$ for $^{132}$Sn
has only a weak correlation with
the neutron skin thickness defined by
$\Delta r_{np} \equiv \sqrt{\langle r^2 \rangle_n} -
\sqrt{\langle r^2 \rangle_p}$ of $^{208}$Pb.
In Fig.~\ref{fig: Pb-Sn},
the $S_\mathrm{PDR}$ for $^{132}$Sn
is shown as a function of the neutron skin 
thickness, $\Delta r_{np}$, of $^{208}$Pb.
The plotted 21 points are obtained by
calculating $\Delta r_{np}$ and $S_\mathrm{PDR}$ with
the SkM$^\ast$ functional, and with
slightly modified values of 10 Skyrme parameters
 ($t_{0,1,2,3}$, $x_{0,1,2,3}$, $W_0$, and $\alpha$).
It seems to indicate some correlation, however, the calculated points
are somewhat scattered.

Using these 21 sample values ($N_d=21$),
the correlation coefficient $r$ is calculated according to Eq. (\ref{r}).
In the present case of Fig. \ref{fig: Pb-Sn},
we obtain the coefficient $r=0.55$.
The correlations between $\Delta r_{np}$ in $^{208}$Pb and
$S_\mathrm{PDR}$ in $^{68}$Ni and $^{78}$Ni,
are also weak with $r= 0.5-0.6$.
Thus, the PDR strength in these spherical (magic) nuclei
indicate a positive correlation
with the skin thickness in $^{208}$Pb,
however, the correlation is weak.
This is qualitatively consistent with the result in Ref. \cite{Reinhard}.

\begin{figure}[tb]
\begin{center}
\includegraphics[width=0.4990\textwidth,keepaspectratio]{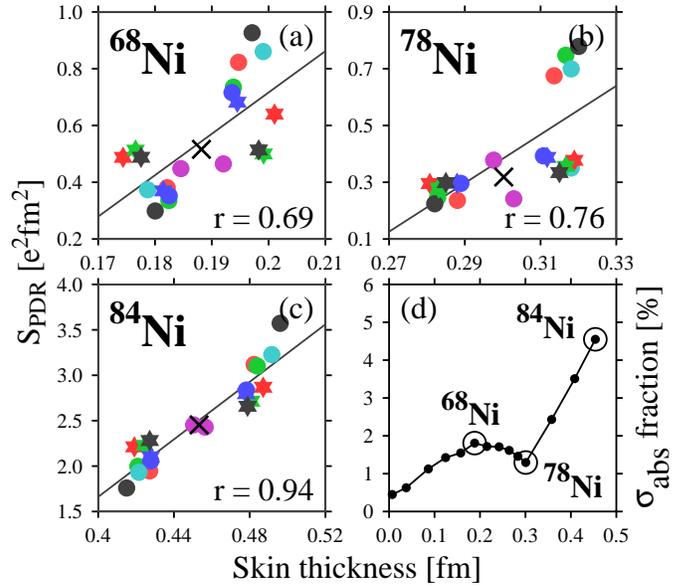}
\caption{(Color online) 
(a)-(c) Correlations between $S_\mathrm{PDR}$
and $\Delta r_{np}$ in $^{68,78,84}$Ni.
See the caption of Fig.~\ref{fig: Pb-Sn}.
Calculated correlation coefficients are also shown.
(d) $f_{\mbox{\sc pdr}}$ as functions of $\Delta r_{np}$ 
for even-even Ni isotopes, calculated with the SkM$^\ast$ parameter set.
See the text for detail.
}
\label{fig: Ni}
\end{center}
\end{figure}
{\it Correlation between $S_\mathrm{PDR}$ and $\Delta r_{np}$ ---}
Next, we discuss the same correlation,
but between the $\Delta r_{np}$ and $S_\mathrm{PDR}$
in the same nucleus.
In Fig. \ref{fig: Ni}, we show the results 
for $^{68}$Ni ($N=40$),
$^{78}$Ni ($N=50$), and $^{84}$Ni ($N=56$).
The scattered data points in Fig. \ref{fig: Ni} (a) suggest
a relatively weak correlation in $^{68}$Ni, 
while the correlation becomes moderately strong for $^{78}$Ni.
The calculated correlation coefficients are $r=0.69$ and 0.76 for
$^{68,78}$Ni, respectively.
In contrast, a very strong linear correlation with $r=0.94$ for $^{84}$Ni
is observed in Fig. \ref{fig: Ni} (c).
It is apparent that
the linear correlation is qualitatively different among the isotopes.

The qualitative difference in $S_\mathrm{PDR}$ 
among the isotopes was previously observed in
the PDR photoabsorption cross section \cite{Inakura11}.
In Ref. \cite{Inakura11},
we systematically calculated, for even-even nuclei up to $Z= 40$,
the PDR fraction $f_\mathrm{PDR}$.
Then, we found that $f_{\mbox{\sc pdr}}$ significantly increases
as a function of neutron number in regions
where the neutron Fermi levels are located
at the weakly-bound low-$\ell$ shells, such as $s$, $p$, and $d$ orbits.
In Ni isotopes, this corresponds to the region with neutron number beyond 50,
as illustrated in Fig. \ref{fig: Ni} (d).
Thus, the present result (Fig. \ref{fig: Ni} (a)-(c)) indicates that
the neutron shell effect also has a significant impact on the linear
correlation between the neutron skin thickness and the PDR strength.

\begin{figure}[tb]
\begin{center}
\includegraphics[width=0.450\textwidth,keepaspectratio]{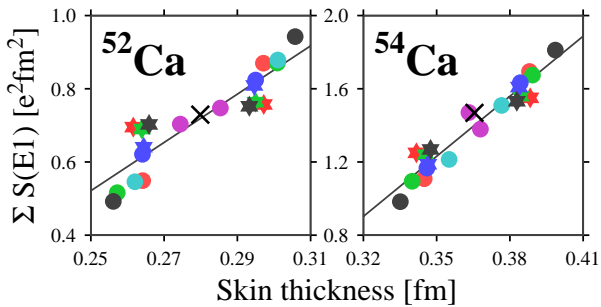}
\caption{{\small (Color online) Same as Fig. \ref{fig: Ni}
but for $^{52,54}$Ca.}}
\label{fig: Ca}
\end{center}
\end{figure}

We confirm the same neutron shell effect
in other light spherical isotopes; $^A$O and $^A$Ca.
For Ca isotopes, the PDR strength appears beyond $N=28$ \cite{Inakura11}.
Accordingly, the strong linear correlation can be seen for
$^{52}$Ca and $^{54}$Ca, in Fig. \ref{fig: Ca}.
The calculated correlation coefficients are
$r=0.91$ and 0.96 for $^{52,54}$Ca, respectively.
These nuclei have neutrons more than 28 and the neutron Fermi level is
located at the $p$ shell.
They are predicted to have the PDR peaks around $E=8$ MeV with
$f_{\mbox{\sc pdr}}\approx 0.03-0.04$ \cite{Inakura11}.
In contrast, nuclei with $N\leq 28$ have very small values of
$f_{\mbox{\sc pdr}}<0.01$ and
the linear correlation in $^{48}$Ca ($N=28$) indicates $r=0.78$
which is much weaker than $^{52,54}$Ca.
For O isotopes, because of the neutron occupation of the $2s$ orbit,
$^{24}$O ($N=16$) provides another example to
show a significant jump in $f_{\mbox{\sc pdr}}$
from $^{22}$O \cite{Inakura11}.
This nucleus has the strongest linear correlation with $r=0.97$.

We also calculate the correlation coefficient for $^{132}$Sn.
It indicates
a relatively weak correlation with $r=0.68$.
Note that $^{132}$Sn corresponds to a kink point
similar to $^{78}$Ni in Fig. \ref{fig: Ni}.
Namely, the PDR fraction in Sn isotopes will jump up beyond $N=82$
\cite{Ebata12}.
The correlation coefficients are summarized
in the second column of Table~\ref{tab: correlation_coef}
for various nuclei.

\begin{table*}
\caption{Calculated correlation coefficients $r$ between
$\Delta r_{np}$ and $S_\mathrm{PDR}$ for selected nuclei.
The SkM$^\ast$ parameter set is adopted as the central values.
The values of variable $\omega_c$ are also listed.
Note that we cannot identify a prominent PDR peak for $^{48}$Ca.
$r^{(v)}$ are obtained with
the variable cutoff energies $\omega_c$ in the fourth row.
The correlation coefficients larger than 0.9 are shown in boldface.
}
\begin{ruledtabular}
\begin{tabular}{l|r|r|rrr|rrr|r|r}
          & $^{24}$O  & $^{26}$Ne & $^{48}$Ca & $^{52}$Ca & $^{54}$Ca
          & $^{68}$Ni & $^{78}$Ni & $^{84}$Ni 
          & $^{58}$Cr & $^{110}$Zr \\
\hline
$r$
&
{\bf 0.97} & 0.83 & 0.78 & {\bf 0.91} & {\bf 0.96} & 0.69 & 0.76 & {\bf 0.94} 
 & 0.80 & 0.74 \\
$r^{(v)}$ 
&
{\bf 0.97} & 0.88 & - & {\bf 0.92} & {\bf 0.94} & 0.77 & {\bf 0.92} & {\bf 0.96} 
 & 0.80 & 0.84 \\
$\omega_c$ [ MeV ]
&
8.29 & 9.95 & - & 10.49 & 9.41 & 11.48 & 8.73 & 8.59 
 & 9.82 & 8.36 \\
\end{tabular}
\end{ruledtabular}
\label{tab: correlation_coef}
\end{table*}

\begin{figure}[tb]
\begin{center}
\includegraphics[width=0.450\textwidth,keepaspectratio]{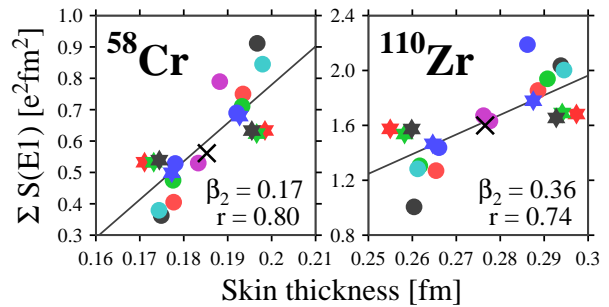}
\caption{(Color online) Same as Fig. \ref{fig: Ni}
but for deformed nuclei $^{58}$Cr and $^{110}$Zr. 
See text for details.}
\label{PDRcorr4}
\end{center}
\end{figure}
{\it Deformed nuclei ---}
The deformation effect seems to somewhat weaken the correlation.
Figure \ref{PDRcorr4} shows two deformed nuclei, $^{58}$Cr with
the quadrupole deformation of $\beta_2=0.17$ and
$^{110}$Zr with a larger deformation of $\beta_2=0.36$.
The $^{58}$Cr nucleus has the same number of neutrons as $^{54}$Ca
carrying a comparable PDR strengths to $^{52}$Ca \cite{Inakura11}.
Nevertheless, the correlation in $^{58}$Cr, $r=0.80$, is significantly weaker
than that in spherical $^{52,54}$Ca.
$^{110}$Zr has an even larger deformation and a weaker correlation,
$r=$0.74, although it has sizable PDR strength. 
The ground-state deformation is expected to produce
a peak splitting both in the PDR and GDR.
Due to the complicated characters in the $E1$ strength distribution,
the PDR strength $S_\mathrm{PDR}$
may be contaminated by the low-energy tail of GDR strength.

\begin{figure}[tb]
\begin{center}
\includegraphics[width=0.450\textwidth,keepaspectratio]{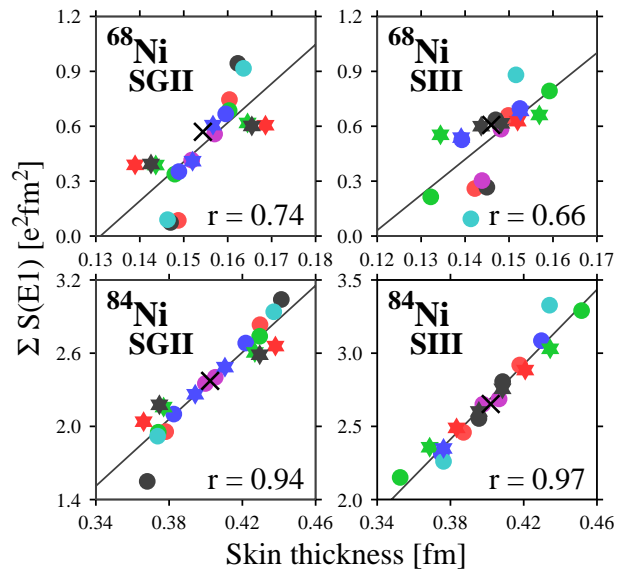}
\caption{(Color online) Same as Fig. \ref{fig: Ni} but for $^{68,84}$Ni with 
SGII and SIII interactions.}
\label{fig: Ni2}
\end{center}
\end{figure}
{\it Universal behaviors ---}
The property of the linear correlation is very robust with respect to
choice of the Skyrme energy functionals.
In Fig. \ref{fig: Ni2},
we show the same correlation plot as Fig. \ref{fig: Ni} calculated with
the parameter set of SkM$^\ast$ replaced by SIII \cite{SIII}
and SGII \cite{SGII}.
All the three 
Skyrme functionals yield a relatively weak correlation for $^{68}$Ni
with $r=0.65-0.75$ and a strong linear correlation for $^{84}$Ni with
$r>0.94$. 
The strong correlation with $r\approx 0.95$ is also confirmed for $^{24}$O and 
$^{54}$Ca.

The slope of the straight line, obtained by linear fit, turns out to be
universal too, with respect to different Skyrme energy functionals.
All these three parameter sets (SkM*, SGII, and SIII)
produce the similar slope,
$d S_\mathrm{PDR}/d(\Delta r_{np})=13-16$
$\mathrm{e}^2\mathrm{fm}$
for $^{84}$Ni.
We observe the linear correlation of
$f_{\mbox{\sc pdr}}$ instead of $S_\mathrm{PDR}$, as well,
with respect to $\Delta r_{np}$.
However, in this case,
the slope obtained by the linear fit has a sizable dependence on
functionals.

{\it Correlation among different energy functionals ---}
Instead of slightly modifying the Skyrme parameters,
we next examine the correlation adopting many different Skyrme functionals
corresponding to a variety of values of the $L$ parameter;
SIII, SGII, SkM$^*$, SLy4 \cite{SLy4}, SkT4 \cite{SkT4},
SkI2, SkI3, SkI4, SkI5 \cite{SkI}, UNEDF0, and UNEDF1 \cite{UNEDF}.
From these eleven different parameter sets, we estimate the
correlation coefficient $r$ in Eq. (\ref{r}) with $N_d=11$.
Again, we have found
a weak correlation with $r=0.47$ for $^{68}$Ni,
and a strong correlation $r=0.89$ for $^{84}$Ni.

We also examine the correlation between the slope parameter of
the symmetry energy $L$ and the PDR fraction $f_\mathrm{PDR}$,
in $^{68}$Ni and $^{84}$Ni.
This leads to the similar coefficients, $r=0.37$ and $0.84$ for
$^{68}$Ni and $^{84}$Ni, respectively.
Thus, to quantitatively constrain $\Delta r_{np}$ and $L$,
the measurement of the PDR in the very neutron-rich $^{84}$Ni
is more favored than in $^{68}$Ni.

The small correlation coefficient between $L$ and $f_\mathrm{PDR}$
for $^{68}$Ni ($r=0.37$) turns out to be
due to the fact that the choice of $\omega_c=10$ MeV has
different meaning for different functionals.
Namely, the different energy functionals produce different PDR peak energies,
some of which are below 10 MeV but some are above that.
The tail of the GDR strength also depends on the choice of
the energy functionals.
Therefore, to make a more sensible analysis for this study,
we should use the variable cutoff $\omega_c$.
This will be discussed below.

\begin{figure}[tb]
\begin{center}
\includegraphics[width=0.30\textwidth,keepaspectratio]{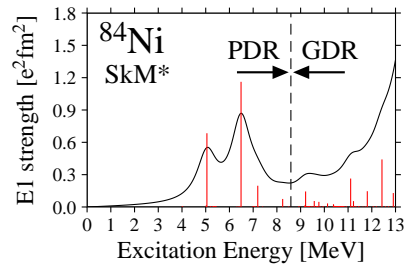}
\caption{(Color online)
Calculated $E1$ strengths ($B(E1;n)$, vertical lines) for $^{84}$Ni
in units of $e^2$fm$^2$
and those smeared with the width of $\gamma=1$ MeV ($S(E1;E)$, solid curve)
in units of $e^2$fm$^2$/MeV.
According to the procedure described in the text,
the cutoff energy is determined as $\omega_c=8.59$ MeV.
}
\label{fig: PDR-GDR}
\end{center}
\end{figure}

{\it Use of variable $\omega_c$ ---}
The PDR strength (\ref{SPDR}) and PDR fraction (\ref{fPDR})
based on variable $\omega_c$ are
hereafter referred to as $S_\mathrm{PDR}^{(v)}$ and
$f_{\mbox{\sc pdr}}^{(v)}$, respectively.
The variable $\omega_c$ is
determined according to the following procedure:
The calculated (discrete) $B(E1)$ values are smeared with the Lorentzian
with the width of $\gamma=1$ MeV.
Plotting this smeared $E1$ strength $S(E1;E)$ as a function of energy,
if we can find a distinguishable PDR peak and its energy $E_{\rm peak}$,
$\omega_c$ is defined as
the energy corresponding to the minimum value of $S(E1;E)$
at $E>E_{\rm peak}$.
In Fig. \ref{fig: PDR-GDR}, as an example,
the determination of $\omega_c$ is shown for $^{84}$Ni.
Since the determination of the variable $\omega_c$ requires
a noticeable PDR peak structure,
it is difficult to define $S_\mathrm{PDR}^{(v)}$ for
most of stable isotopes.

The values of $\omega_c$ varies from nucleus to nucleus
within a range of $10\pm 2$ MeV for those
listed in Table~\ref{tab: correlation_coef}.
Note that $\omega_c$ may also change when we slightly modify the Skyrme
parameters.
Although the correlation is slightly enhanced by replacing
$S_\mathrm{PDR}$ by $S_\mathrm{PDR}^{(v)}$ in most cases,
they are approximately similar, $r^{(v)}\approx r$.
In Table~\ref{tab: correlation_coef}, there are a few exceptions;
$^{78}$Ni ($r=0.76\rightarrow r^{(v)}=0.92$),
$^{68}$Ni ($r=0.69 \rightarrow r^{(v)}=0.77$),
and deformed $^{110}$Zr ($r=0.74 \rightarrow r^{(v)}=0.84$).
In these cases, we found that the separation between PDR and GDR is
somewhat ambiguous, and the results depend on the choice of $\omega_c$.
On the other hand,
isotopes indicating $r>0.9$ with fixed $\omega_c=10$ MeV
show $r^{(v)}\approx 1$ with variable $\omega_c$ as well.
In Ni isotopes,
although the value of $r^{(v)}$ are slightly different from $r$,
it is confirmed that the linear correlation is significantly stronger in
$^{84}$Ni than in $^{68}$Ni.

\begin{figure}[tb]
\begin{center}
\includegraphics[width=0.450\textwidth,keepaspectratio]{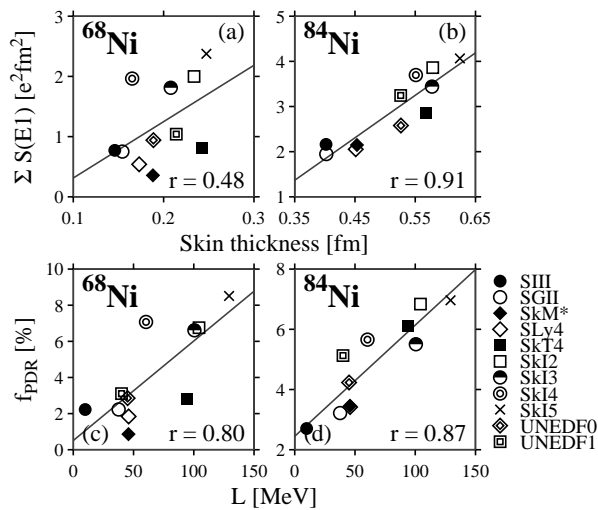}
\caption{Correlations between $S_\mathrm{PDR}^{(v)}$
and $\Delta r_{np}$ (top panels),
and between $f_{\mbox{\sc pdr}}^{(v)}$ and $L$ (bottom)
for $^{68}$Ni (left) and $^{84}$Ni (right),
among eleven different Skyrme functionals.
}
\label{fig: functional}
\end{center}
\end{figure}

For eleven different parameter sets,
the correlation between $S_\mathrm{PDR}^{(v)}$ and $\Delta r_{np}$
for $^{68,84}$Ni is shown in the upper part of Fig.~\ref{fig: functional}.
A strong positive correlation ($r^{(v)}>0.9$) between the PDR strength
$S_\mathrm{PDR}^{(v)}$ and $\Delta r_{np}$ can be seen in $^{84}$Ni. 
In contrast, it is significantly weaker for $^{68}$Ni ($r^{(v)}=0.48$).
The bottom part of Fig.~\ref{fig: functional} shows correlation
between $f_\mathrm{PDR}^{(v)}$ and the slope parameter $L$ of the symmetry energy.
Again, the correlation is stronger for $^{84}$Ni with $r^{(v)}=0.87$
than $^{68}$Ni with $r^{(v)}=0.80$. 
The correlation between $\Delta r_{np}$ and $L$ has similar trend, $r^{(v)}=0.88$ 
for $^{84}$Ni and $r^{(v)}=0.84$ for $^{68}$Ni. 
Basic features of the correlation with the variable $\omega_c$ are consistent
with those obtained with $\omega_c$ fixed at 10 MeV.
Thus, the PDR strength in $^{84}$Ni with many excess neutrons
can provide a better constraint on $L$ and the neutron skin,
compared to $^{68}$Ni.

{\it Summary ---}
We have studied the correlation of the PDR and the neutron skin 
thickness, for nuclei with $Z\leq 40$ and $^{132}$Sn.
We have found that the strong linear correlation is seen only in particular
nuclei.
The PDR strength has a very strong linear correlation with the neutron 
skin thickness in spherical neutron-rich nuclei with $14<N\leq 16$,
$28 < N \leq 34$, and $50 < N \leq 56$.
In these regions, the neutron Fermi levels are located at the loosely-bound 
low-$\ell$ shells and the PDR strengths significantly increase as
the neutron number.
Nuclei outside of these regions have weaker correlations. 
This linear correlation is robust with respect to the choice of the energy 
functional parameter set.
This suggests that the experimental observation of PDR
in properly selected neutron-rich nuclei
could be a possible probe of the neutron skin thickness $\Delta r_{np}$
and a constraint on
the slope parameter $L$ of the symmetry energy.
The linear correlation seems to be weakened by the deformation due to 
the peak splitting of the PDR and the GDR.
The present result may provide a solution for the controversial issue on
the correlation between the PDR and the neutron skin,
for which different conclusions were reported previously
\cite{Volz,Klimkiewicz,Piekarewicz06,Reinhard}.

\section*{Acknowledgments}

This work is partly supported by HPCI System Research Projects
(Project ID: hp120192 and hp120287),
by Collaborative Interdisciplinary Program
(Project ID: 13a-33, 12a-20 and 11a-21) at University of Tsukuba,
and by JSPS KAKENHI Grant numbers
21340073, 24105006, 25287065, and 25287066.
The numerical calculations were partially performed on 
the RIKEN Integrated Cluster of Clusters (RICC) as well.


\begin{thebibliography}{99}
\bibitem{Ste05}
A. W. Steiner et al., Phys. Rep. {\bf 411}, 325 (2005).

\bibitem{Li08}.
B. A. Li et al., Phys. Rep. {\bf 464}, 113 (2008).

\bibitem{Bro00} 
B. A. Brown, Phys. Rev. Lett. {\bf 85}, 5296 (2000).

\bibitem{Fur02} 
R. J. Furnstahl, Nucl. Phys. A {\bf 706}, 85 (2002).

\bibitem{Klimkiewicz} 
A. Klimkiewicz et al., Phys. Rev. {\bf C 76}, 051603(R) (2007).

\bibitem{Volz} 
S. Volz et al., Nucl. Phys. {\bf A 779}, 1 (2006).

\bibitem{Ebata10} 
S. Ebata et al.,
Phys. Rev. {\bf C 82}, 034306 (2010). 

\bibitem{Piekarewicz06}
J. Piekarewicz, Phys. Rev. {\bf C 73}, 044325 (2006),

\bibitem{Carbone} 
A. Carbone et al.,
Phys. Rev. {\bf C 81}, 041301(R) (2010).

\bibitem{Reinhard} 
P.-G. Reinhard and W. Nazarewicz, Phys. Rev. {\bf C 81}, 051303(R) (2010).

\bibitem{Reinhard2} 
P.-G. Reinhard and W. Nazarewicz, Phys. Rev. {\bf C 87}, 014324 (2013).

\bibitem{Inakura11}
T. Inakura, T. Nakatsukasa, and K. Yabana, Phys. Rev. {\bf C 84}, 021302(R) (2011).

\bibitem{Nakatsukasa07} 
T. Nakatsukasa, T. Inakura, and K. Yabana, Phys. Rev. {\bf C 76}, 024318 (2007).

\bibitem{Inakura09} 
T. Inakura, T. Nakatsukasa, and K. Yabana, Phys. Rev. {\bf C 80}, 044301 (2009).

\bibitem{Avogadro11}
P. Avogadro and T. Nakatsukasa, Phys. Rev. {\bf C 84}, 014314 (2011).

\bibitem{Stoitsov11}
M. Stoitsov, M. Kortelainen, T. Nakatsukasa, C. Losa, and W. Nazarewicz,
Phys. Rev. {\bf C 84}, 041305(R) (2011).

\bibitem{Avogadro13}
P. Avogadro and T. Nakatsukasa, Phys. Rev. {\bf C 87}, 014331 (2013).

\bibitem{Inakura06} 
T. Inakura et al.,
Nucl. Phys. A 768, 61 (2006).

\bibitem{Nakatsukasa05} 
T. Nakatsukasa and K. Yabana, Phys. Rev. {\bf C 71}, 024301 (2005).

\bibitem{SkM*} 
J. Bartel, P. Quentin, M. Brack, C. Guet, and H.B. H\r{a}kansson, Nucl. Phys. {\bf A 386} (1982) 79.

\bibitem{Ebata12} 
S. Ebata, T. Nakatsukasa, and T. Inakura, 
arXiv: 1111.0362; arXiv: 1201.3462.

\bibitem{SIII} 
M. Beiner, H. Flocard, Nguyen van Giai, and P. Quentin, Nucl. Phys. {\bf A 238} (1975) 29.

\bibitem{SGII} 
Nguyen Van Giai, and H. Sagawa, Phys. Lett. {\bf B 106}, 379 (1981).

\bibitem{Piekarewicz12}
J. Piekarewicz et al.,
 Phys. Rev. C {\bf 85}, 041302 (2012).

\bibitem{SLy4} 
E. Chabanat, P. Bonche, P. Haensel, J. Mayer, and R. Schaeffer, Nucl. Phys. {\bf A 627} (1998) 231.

\bibitem{SkT4} 
F. Tondeur, M. Brack, M. Farine, and J.M. Pearson, Nucl. Phys. {\bf A 420} (1984) 297.

\bibitem{SkI} 
P.-G. Reinhard and H. Flocard, Nucl. Phys. {\bf A 584}, (1995) 467.

\bibitem{UNEDF} 
M. Kortelainen et al.,
Phys. Rev. {\bf C 85}, 024304 (2012)

\end{thebibliography}
\end{document}